\documentclass[article]{IEEEtran}

\usepackage{blindtext, graphicx}
\usepackage{balance}
\usepackage{tikz}
\usepackage{listings}

\usepackage[font=small]{caption}
\usepackage[left=19.1mm, right=19.1mm, top=25.4mm, bottom=19.11mm]{geometry}

\usepackage[utf8]{inputenc}

\begin{document}

!TeX TS-program = pdflatex

\title{Low-Frequency Load Identification using CNN-BiLSTM Attention Mechanism}

%\author{ \parbox{3 in}{\centering Huibert Kwakernaak
%         \thanks{*Use the $\backslash$thanks command to put information here}\\
%         Faculty of Electrical Engineering, Mathematics and Computer Science\\
%         University of Twente\\
%         7500 AE Enschede, The Netherlands\\
%         {\tt\small h.kwakernaak@autsubmit.com}}
%         \hspace*{ 0.5 in}
%         \parbox{3 in}{ \centering Pradeep Misra**
%         \thanks{**The footnote marks may be inserted manually}\\
%        Department of Electrical Engineering \\
%         Wright State University\\
%         Dayton, OH 45435, USA\\
%         {\tt\small pmisra@cs.wright.edu}}
%}

% author names and affiliations
% use a multiple column layout for up to three different
% affiliations

\author{Amanie Azzam, Saba Sanami, and Amir G. Aghdam}
\thanks{Amanie Azzam, Saba Sanami and Amir G. Aghdam are with the Department of Electrical and Computer Engineering, Concordia Univerity, Montreal, QC, Canada. Email: {\tt\small  amanie.azzam@mail.concordia.ca, saba.sanami@mail.concordia.ca, amir.aghdam@concordia.ca}}
\thanks{This work has been supported by the Natural Sciences and Engineering Research Council of Canada (NSERC) under grant RGPIN-262127-17.}

%\doublespacing

\maketitle
\thispagestyle{empty}
\pagestyle{empty}

%%%%%%%%%%%%%%%%%%%%%%%%%%%%%%%%%%%%%%%%%%%%%%%%%%%%%%%%%%%%%%%%%%%%%%%%%%%%%%%%
\begin{abstract}

Non-intrusive Load Monitoring (NILM) is an established technique for effective and cost-efficient electricity consumption management. The method is used to estimate appliance-level power consumption from aggregated power measurements. This paper presents a hybrid learning aproach, consisting of convolutional neural network (CNN) and a  bidirectional long short-term memory (BILSTM), featuring an integrated attention mechanism, all within the context of disaggregating low-frequency power data. While prior research has been mainly focused on high-frequency data disaggregation, our study takes a distinct direction by concentrating on low-frequency data. The proposed hybrid CNN-BILSTM model is adept at extracting both temporal (time-related) and spatial (location-related) features, allowing it to precisely identify energy consumption patterns at the appliance level. This accuracy is further enhanced by the attention mechanism, which aids the model in pinpointing crucial parts of the data for more precise event detection and load disaggregation. We conduct simulations using the existing low-frequency REDD dataset to assess our model’s performance. The results demonstrate that our proposed approach outperforms existing methods in terms of accuracy and computation time.


\end{abstract}

%%%%%%%%%%%%%%%%%%%%%%%%%%%%%%%%%%%%%%%%%%%%%%%%%%%%%%%%%%%%%%%%%%%%%%%%%%%%%%%%
\section{Introduction}

Non-intrusive load monitoring (NILM) has emerged as a critical technology in the field of energy consumption analysis and management. It enables one to quantify individual home appliances' energy consumption without requiring additional instrumentation or sensor attachments. As the society continues to grapple with the challenges of rising energy demands and environmental concerns, NILM stands as a promising solution to promote energy efficiency, reduce wastage, and enable informed decision-making for homeowners and utilities alike.

The benefits of NILM extend to both consumers and energy providers. For consumers, NILM empowers homeowners with insights into their energy consumption habits. This awareness encourages energy-efficient behaviour, resulting in reduced utility bills and a lower carbon footprint. It also provides real-time information about appliance operation, helping users detect malfunctioning or inefficient devices \cite{ref1}. This, in turn, facilitates timely maintenance and replacement decisions. Integrating NILM with smart home systems allows for intelligent energy management \cite{ref1-1, ref1-2}. Appliances can be automatically controlled based on user preferences and real-time energy consumption data. For energy providers and utilities, NILM facilitates demand response programs by enabling utilities to communicate with consumers and control appliances during peak demand periods, thereby balancing the grid's load. Accurate appliance-level consumption data aids in load forecasting and peak demand management \cite{ref1-3}. This enables utilities to allocate resources effectively and prevent grid overloads \cite{ref1-4}. Detailed appliance-level data also helps utilities strategically plan infrastructure upgrades and expansions, improving grid reliability and efficiency.

NILM was first introduced by Hart \cite{ref1-5} during the early 1900s. The concept encompasses four distinct components: event detection, data processing, load decomposition, and load identification. Load identification is a fundamental task in NILM among these components. It consists of sophisticated algorithms which categorize loads using input features either provided by human input or extracted from electrical signal attributes. The early investigations into load identification effectiveness \cite{ref2, ref3, ref4} often faced constraints arising from emerging hardware and software capacities.  

Traditional NILM methods focused on statistical analysis and pattern recognition \cite{ref1-5, ref6}. Rule-based methods and statistical analyses constitute prevalent non-learning techniques, which rely on predefined heuristics and pattern recognition algorithms like hidden markov models (HMMs) \cite{ref1-5}, and clustering methods \cite{ref7}. These techniques often struggle to capture the complicated variability and complexities of modern appliances' behaviours, particularly in cases of overlapping power signatures or subtle differences.

The introduction of deep learning marked a turning point in load disaggregation research, overcoming many of the shortcomings of non-learning-based methods \cite{ref7, ref8, ref9}. The preliminary utilization of deep neural networks for tackling NILM challenges was reported in \cite{ref11}. This seminal work introduced a composite architecture combining LSTM networks and denoising autoencoders (DAE) to construct a rectangular network. Additionally, Davies et al. devised an automatic feature learning approach in \cite{ref12}, wherein high-frequency electrical data was downsampled into four channels. This data transformation facilitated the application of a five-layer CNN to achieve effective classification. Furthermore, a fusion of sequence-to-point, spatial, and channel attention mechanisms were detailed in \cite{ref13}. The outcome was the formulation of a convolutional block attention model, designed to comprehensively learn the distinctive characteristics of target devices.

In this research, it is desired to utilize low-frequency power data to develop a NILM algorithm. The proposed approach offers a cost-effective advantage, as low-frequency data collection requires less expensive measurement equipment compared to its high-frequency counterpart. Our methodology entails the application of a CNN-BILSTM model with an attention mechanism. The CNN-BILSTM model combines two distinct neural network architectures, to leverage the strengths of each for enhanced NILM performance. The CNN part specializes in detecting spatial features and patterns, which is crucial for distinguishing different appliances based on their distinct energy consumption signatures. The BILSTM component, on the other hand, excels at capturing temporal patterns and dependencies within sequential data, making it adept for load disaggregation tasks. The integration of an attention mechanism further refines the model's focus on relevant information, improving its ability to accurately identify appliance activation and deactivation.

The rest of the paper is organized as follows: Section~II defines the problem, followed by Section~III, which outlines the CNN-BILSTM network combined with the attention mechanism. Comparative simulation results are presented in Section~IV, along with the corresponding accuracy metrics. Finally, concluding remarks are given in Section~V.

\section{Problem Statement}

Consider the energy consumption data for a household, available as an aggregate signal. It is desired to decompose the overall aggregated power waveform into its constituent appliance-level components as depicted in Fig.~1. The main challenges for power disaggregation include: (i) handling appliances' power signature variations \cite{ref14}; (ii) identifying appliances whose power signature overlap partially or completely \cite{ref16}, and (iii) achieving the desired accuracy. Previous research efforts have tackled these issues by utilizing high-frequency data, as they contain the complete signal and can be used to extract maximum information. However, the significant drawback lies in the cost of gathering such high-frequency data. It demands dedicated infrastructure, which incurs costs in terms of installing hardware to collect this type of data. Conversely, low-frequency features come with information loss but offer the advantage of being easily collectible \cite{ref17-1}. The forward challenge revolves around the disaggregation of power consumption data, characterized by a low frequency of 0.1 Hz. 
Mathematically, the disaggregation is formulated in \cite{ref1-5} as follows:
$$
P_{total}(t)=\sum_{i=1}^N P_{{appliance}_i}(t)+P_{noise}(t), \eqno{(1)}
$$where $P_{total}(t)$  represents the total power consumption at time $t$, $P_{{appliance}_i}(t)$ is the power consumption of the $i$-$th$ appliance, $N$ is the number of appliances, and $P_{noise}(t)$ signifies the noise component in the signal. The objective is to estimate $P_{{appliance}_i}(t)$ for each appliance, despite challenges posed by low-frequency measurements and signal noise.

\begin{figure}
    \centering
    \includegraphics[width=0.4\textwidth]{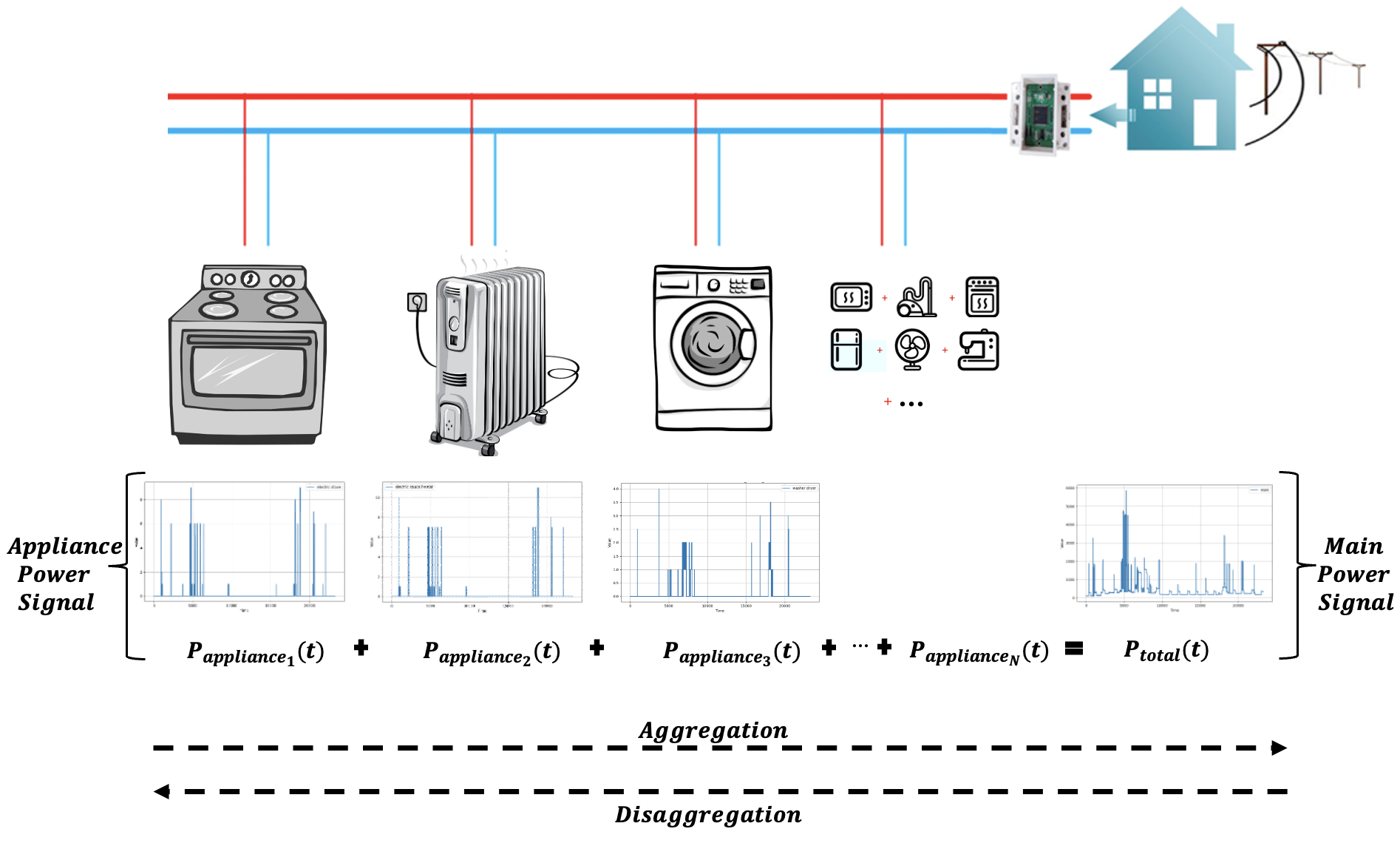}
    \caption{Household appliances power consumption}
    \label{fig:figure1.1}
\end{figure}

\section{Main Results}
The hybrid model is presented in this section.
\subsection{Convolutional Neural Network (CNN)}

CNNs are utilized for the classification of load types, leveraging spatial patterns within the data \cite{ref22}\cite{ref23}. The CNN architecture consists of convolutional layers followed by pooling layers. Each convolutional layer detects progressively more complex features. It can learn to distinguish the energy consumption patterns of various appliances, e.g., the distinct start-up and shut-down sequences of a dishwasher and an electric stove. Pooling layers reduce spatial dimensions, preserving important information. This process is particularly essential for our application, as it helps condense the information while preserving the critical spatial characteristics that distinguish appliances.
The convolution operation utilized in \cite{ref25} is:
$$
C_{ij}= \sum_{m=1}^M\sum_{n=1}^NI_{i+m,j+n} K_{m,n}, \eqno{(9)}
$$where $I$ is the input matrix, $K$ is the convolutional kernel, and $M$ and $N$ are the kernel dimensions.
Pooling (max-pooling) extracts the maximum value from a region of the input, reducing spatial dimensions while retaining significant features denoted by $P_{ij}$, i.e.:
$$
P_{ij} = \max(I_{2i,2j}, I_{2i,2j+1}, I_{2i+1,2j}, I_{2i+1,2j+1}). \eqno{(10)}
$$
A CNN is an essential part of our load identification and appliance classification technique because it is a powerful tool for capturing the spatial patterns in energy consumption data. This enables our model to effectively distinguish between appliances based on their unique energy consumption signatures.

\subsection{Long Short Term Memory (LSTM)}

Long short-term memory (LSTM) is a type of recurrent neural network (RNN) designed to process sequential data and remember long-term dependencies \cite{ref18}. It is widely used in deep learning for sequence prediction tasks such as natural language processing \cite{ref19}, speech recognition \cite{ref20}, and time series forecasting \cite{ref21}. The core idea behind LSTM is the use of specialized memory cells that can store and retrieve information over long sequences. 

\subsection{Bidirectional Long Short Term Memory (BiLSTM)}

The BiLSTM unit plays a crucial role in the proposed disaggregation process. Its bidirectional processing capability helps our model comprehend sequential data and identify the complex temporal features in energy consumption data.
Home appliances have unique energy consumption patterns that can be observed (and learned) over time. These patterns have nuanced transitions and interactions. The BiLSTM is a powerful tool for capturing these subtleties. It consists of two LSTM layers, one processing the input sequence in the forward direction and the other in reverse, enabling it to capture more context (compared to the LSTM) from both past and future inputs. The method can recognize the temporal sequences associated with appliance operation, such as the recurring cycles of a refrigerator’s compressor or the periodic fluctuations in power consumption caused by a washing machine’s agitator.
Mathematically, the forward LSTM computes hidden states as follows \cite{ref24}:
$$
h_t^f= LSTM_f (x_t,h_{t-1}^f,c_{t-1}^f), \eqno{(2)}
$$where $x_t$ is the input at time $t$, $h_{t-1}^f$ is the forward hidden state at time $t-1$, and $c_{t-1}^f$ is the forward cell state at time $t-1$.
The backward LSTM computes hidden states in a similar manner i.e.\cite{ref24}:
$$
 h_t^b= LSTM_b (x_t,h_{t+1}^b,c_{t+1}^b), \eqno{(3)}
$$where $h_{t+1}^b$ is the forward hidden state at time $t+1$, and $c_{t+1}^f$ is the forward cell state at time $t+1$.
The final hidden state is a concatenation of the forward and backward hidden states:
$$
h_t=[h_t^f,h_t^b]. \eqno{(4)}
$$
The BiLSTM is a powerful tool that helps the proposed model distinguish between different appliances based on their temporal intricacies. It can accurately break down total power consumption into individual appliance-level components by effectively modeling temporal patterns and dependencies.

\subsection{Attention Mechanism}

The attention mechanism is integrated into the BiLSTM model to enhance the model's capability to focus on specific time steps crucial for load identification. Attention mechanisms have been widely adopted in sequence-to-sequence tasks, and in our context, they dynamically weigh the importance of different time steps within the input sequence. This dynamic weighting is essential for accurately identifying appliance patterns, as certain time steps may contain more relevant information than others. 
For better identification of appliances, one can also consider the turning on and off time instants, especially in relation to other appliances, as features to improve the accuracy of the method. The model uses the attention mechanism to assign weights to each moment in the input sequence. Therefore, the moments when significant changes occur can be used in the identification task. This leads to a more accurate disaggregation of the power consumption data into the individual components.

The attention score $\alpha_t$ for a time step $t$ is calculated using the hidden state $h_t$ and a context vector $v$ according to \cite{ref26}:
$$
\alpha_t = \frac{\mathrm{exp}(e_t)}{\sum_{j=1}^T \mathrm{exp}(e_j)}. \eqno{(5)}
$$$e_t$ in the above equation is the attention energy at time $t$, computed as:
$$
e_t = v^T \mathrm{tanh}\,(W_h h_t + b_h), \eqno{(6)}
$$where $W_h$ and $b_h$ are weight and bias learnable parameters for attention calculation, and $T$ is the total number of time steps.
The context vector $c$ is obtained as the weighted sum of hidden states according to \cite{ref26}:
$$
c= \sum_{t=1}^T \alpha_t h_t. \eqno{(7)}
$$The final prediction $y$ in the regression task, according to \cite{ref26}, is then obtained by passing $c$ through a linear layer as follows:
$$
y= W_y c+ b_y, \eqno{(8)}
$$where $W_y$ and $b_y$ are weight and bias learnable parameters for linear transformation in regression.

The attention mechanism is a mechanism for improving load identification performance by offering a more focused and selective understanding of the data. This allows the model to better identify important temporal events that define appliance behaviour, ultimately leading to more accurate and reliable load disaggregation. Therefore, integrating an attention mechanism is crucial for improving the model’s ability to recognize patterns of appliance activation and deactivation.

\subsection{Proposed Model}

In our study, we propose a hybrid model for the precise identification of home appliances' energy consumption patterns. Our model seamlessly integrates three key components: a CNNs, a BiLSTM network, and an attention mechanism for pinpointing critical time steps within energy data sequences. The CNNs leverage spatial patterns in the data to categorize the identified appliances effectively. The BiLSTM, on the other hand, captures intricate temporal dependencies in energy consumption data for enabling accurate appliance identification. Finally, the attention mechanism enhances the model to focus on more informative time steps, crucial for accurate load estimation. This hybrid approach, shown in Fig.~2, capitalizes on the strengths of each component to create a robust and comprehensive solution for load identification using low-frequency power data.

\begin{figure}
  \centering
  \includegraphics[width=0.4\textwidth]{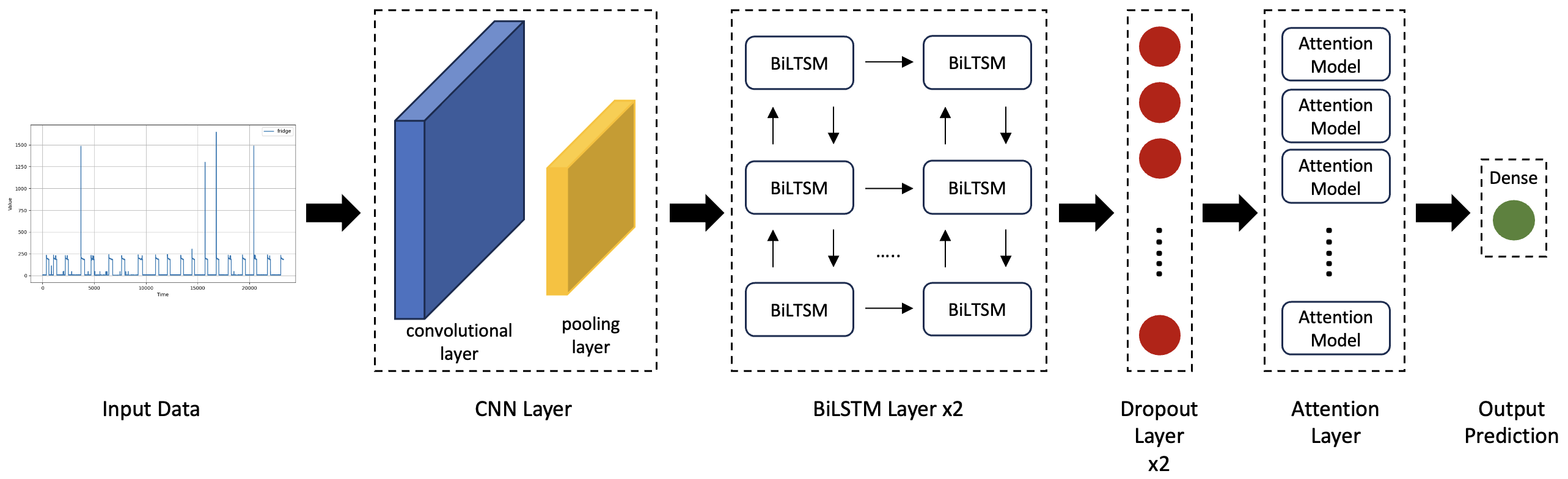}
  \caption{Proposed model architecture}
  \label{fig:example}
\end{figure}

\section{Simulations}

The REDD dataset is a well-established benchmark in the field of load identification and energy disaggregation. This dataset comprises a rich set of electrical load measurements recorded from various sensors and appliances within a residential setting. We will use the data for six appliances: dishwasher, electric space heater, electric stove, refrigerator, microwave, and washer dryer. It is publicly available and was collected by the authors of \cite{ref14}. The dataset is characterized by a low monitoring frequency of 1 Hz. We reduced the sampling frequency of the REDD dataset from 1 Hz to 0.1 Hz by downsampling in time. To prepare the data for our experiments, we perform standard preprocessing steps as part of the proposed model shown in Fig.~3, including data cleaning, normalization, and sequence splitting into training and testing sets. Additionally, we transform the data into a format suitable for both the BiLSTM model (for regression) and the CNN model (for classification).
Table I provides an overview of the proposed model's hyperparameters: epochs, cost function, optimization method, and the configuration of various neural network layers, used in the simulations. It is to be noted that 80\% of data was used for training and 20\% was used for testing. The proposed method was trained over a span of 20 epochs. The normalized data for appliances and total household energy consumption are shown in Fig.~4 and 5 respectively.

The training and validation loss values are shown for each epoch as depicted in Fig. 6. This figure demonstrates that the model successfully converges, with a loss value lower than 0.00025.
Despite the depth and complexity of the proposed model, it is noteworthy that the runtime (computation time) was notably fast (around 31 s 19ms/step). This computational efficiency is essential for real-time or near-real-time applications, where timely load identification is paramount.
The power consumption of the refrigerator, microwave, electric space heater, and electric stove are shown in Fig.~7, 8, 9 and 10, where each appliance is properly identified by the proposed method

\usetikzlibrary{shapes,arrows}
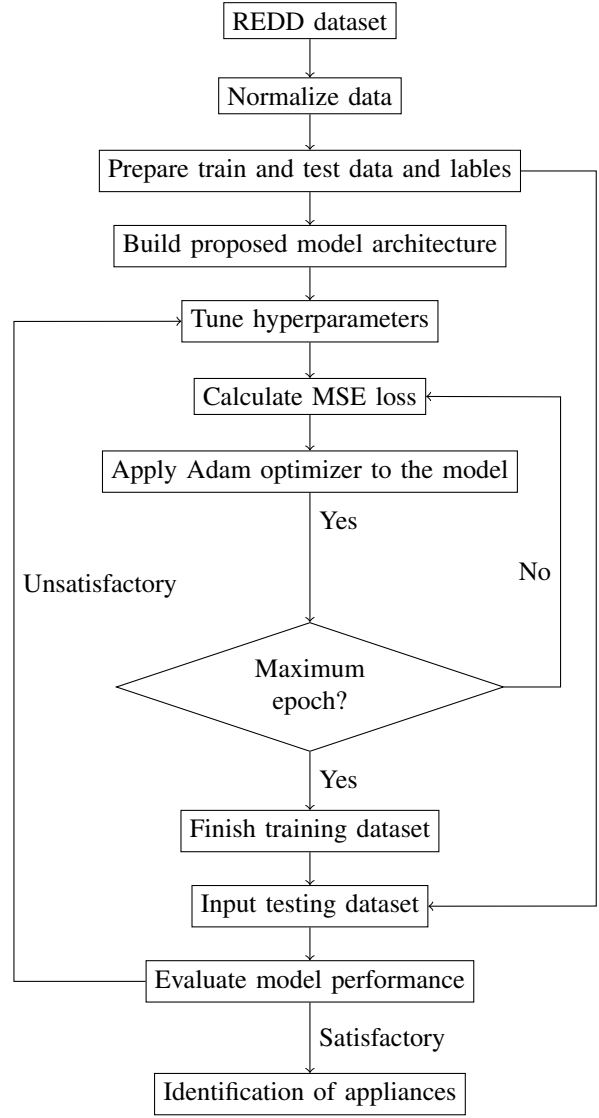
\begin{figure}
    \centering
    \begin{tikzpicture}
    %\node [rectangle, draw] (start) {Start};
    \node [rectangle, draw] (process1) {REDD dataset};
    \node [rectangle, draw, below of=process1] (process2) {Normalize data};
    \node [rectangle, draw, below of=process2] (process3) {Prepare train and test data and lables};
    %\node [rectangle, draw, below of=process3] (process3-1) {Create mini-batches of size 32};
    \node [rectangle, draw, below of=process3] (process4) {Build proposed model architecture};
    \node [rectangle, draw, below of=process4] (process4-1) {Tune hyperparameters};
    \node [rectangle, draw, below of=process4-1] (process5) {Calculate MSE loss};
    %\node [rectangle, draw, below of=process5] (process6) {Train 80\% and test 20\% of data};
    \node [rectangle, draw, below of=process5] (process7) {Apply Adam optimizer to the model};
    %\node [diamond, draw, below=9cm, shape aspect=3, text width=2cm, align=center] (process8) {Finish one epoch?};
    \node [diamond, draw, below=8cm, shape aspect=3, text width=2cm, align=center] (process9) {Maximum epoch?};
    \node [rectangle, draw, below=10.5cm] (process10) {Finish training dataset};
    \node [rectangle, draw, below of=process10] (process11) {Input testing dataset};
    \node [rectangle, draw, below of=process11] (process12) {Evaluate model performance};
    \node [rectangle, draw, below=14cm] (process13) {Identification of appliances};
    %\node [rectangle, draw, below of=process13] (end) {End};
    
    %\draw [->] (start) -- (process1);
    \draw [->] (process1) -- (process2);
    \draw [->] (process2) -- (process3);
    %\draw [->] (process3) -- (process3-1);
    \draw [->] (process3) -- (process4);
    \draw [->] (process4) -- (process4-1);
    \draw [->] (process4-1) -- (process5);
    %\draw [->] (process5) -- (process6);
    \draw [->] (process5) -- (process7);
    %\draw [->] (process7) -- (process9);
     % Arrow from Process 8 to Process 5 with "NO" label
    %\draw [->] (process8.east) -- ++(1cm,0) -- node[midway, left] {No} ++(0,3.87cm) -- (process5.east);
    % Arrow from Process 9 to Process 5 with "NO" label
    \draw [->] (process9.east) -- ++(0.75cm,0) -- node[pos=0.4, left] {No} ++(0,3.86cm) -- (process5.east);
    % Arrow from Process 7 to Process 9 with "Yes" label
    \draw [->] (process7.south) -- node[pos=0.9, right] {Yes} ++(0,-0.4cm) -- (process9.north);
    % Arrow from Process 9 to Process 10 with "Yes" label
    \draw [->] (process9.south) -- node[pos=0.9, right] {Yes} ++(0,-0.4cm) -- (process10);
    \draw [->] (process10) -- (process11);
     % Arrow from Process 3 to Process 11
    \draw [->] (process3.east) -- ++(1.0cm,0) -- ++(0,-9.78cm) -- (process11.east);
    % Arrow from Process 12 to Process 4 with "Bad" label
    \draw [->] (process12.west) -- ++(-1.75cm,0) -- node[pos=0.6, right] {Unsatisfactory} ++(0,8.78cm) -- (process4-1.west);
    \draw [->] (process11) -- (process12);
    \draw [->] (process12) -- node[pos=0.7, right] {Satisfactory} ++(0,-1cm) -- (process13);
    %\draw [->] (process13) -- (end);
    \end{tikzpicture}
 \caption{Flowchart of the proposed model}
\end{figure}

\begin{table}
    \centering
    \caption{The proposed model's hyperparameters used in the simulations}
    \begin{tabular}{|c|c|c|c|}
        \hline
        hyperparameters & method \\
        \hline
        No. of epochs & 20\\
        Loss function & MSE\\
        Optimizer & Adam\\
        No. of CNN layer & 1\\
        No. of BiLSTM layer & 2\\
        No. of Attention mechanism layer & 1\\
        No. of Dropout layer & 2\\
        No. of Dense layer & 1\\
        \hline
    \end{tabular}
    \label{tab:1}
\end{table}

\begin{figure}
  \centering
  \includegraphics[width=0.4\textwidth]{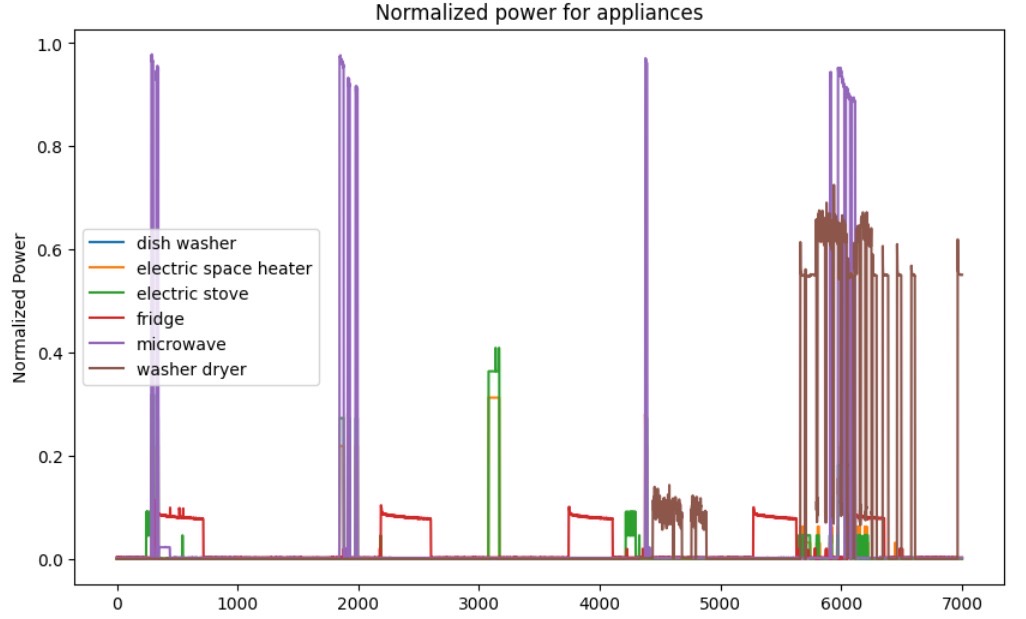}
  \caption{Normalized appliance data}
  \label{fig:example}
\end{figure}
\begin{figure}
  \centering
  \includegraphics[width=0.4\textwidth]{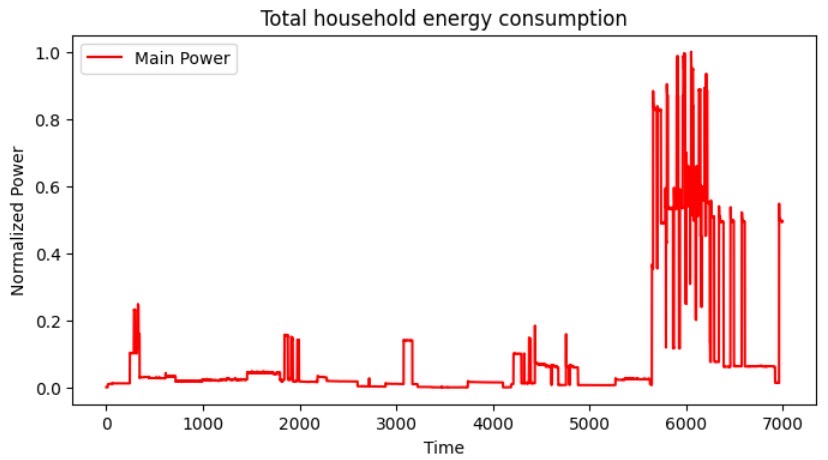}
  \caption{Normalized total household energy consumption data}
  \label{fig:example}
\end{figure}

\begin{figure}
  \centering
  \includegraphics[width=0.4\textwidth]{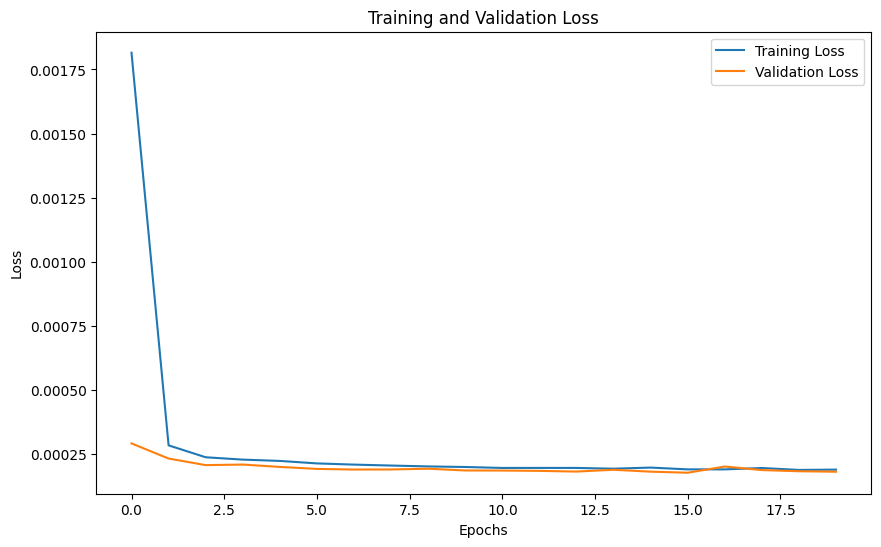}
  \caption{Training and validation loss}
  \label{fig:example}
\end{figure}

\begin{figure}
  \centering
  \includegraphics[width=0.4\textwidth]{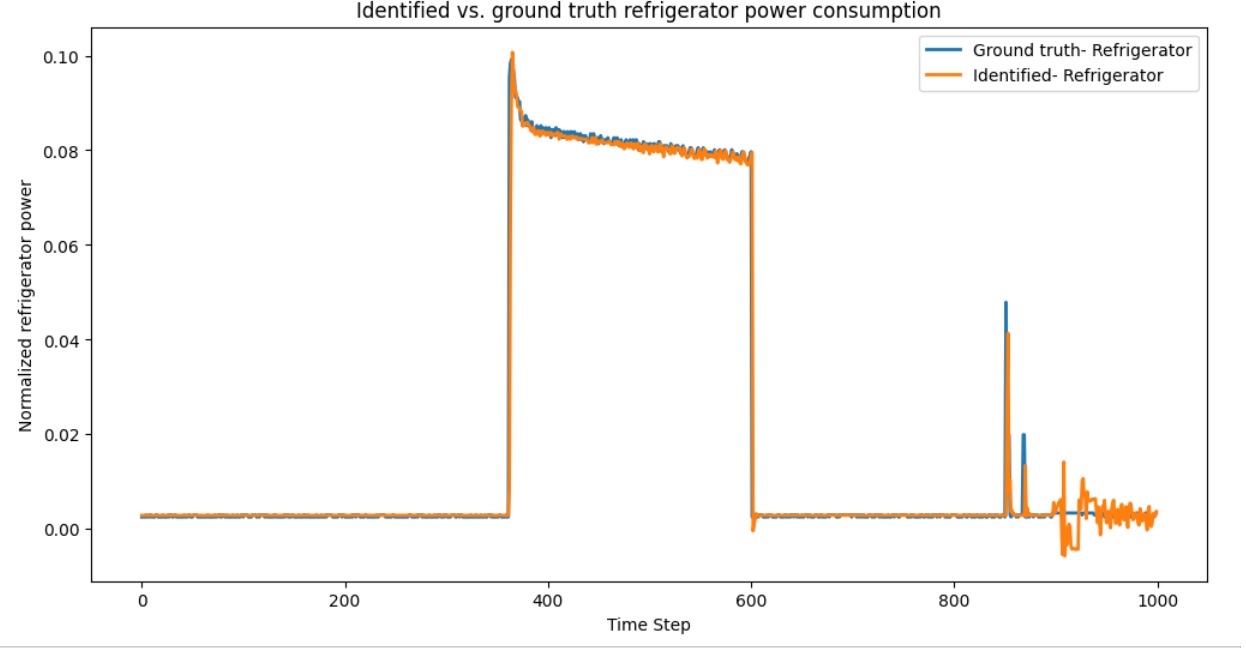}
  \caption{Identified vs. ground truth power consumption for a refrigerator}
  \label{fig:example}
\end{figure}

\begin{figure}
  \centering
  \includegraphics[width=0.4\textwidth]{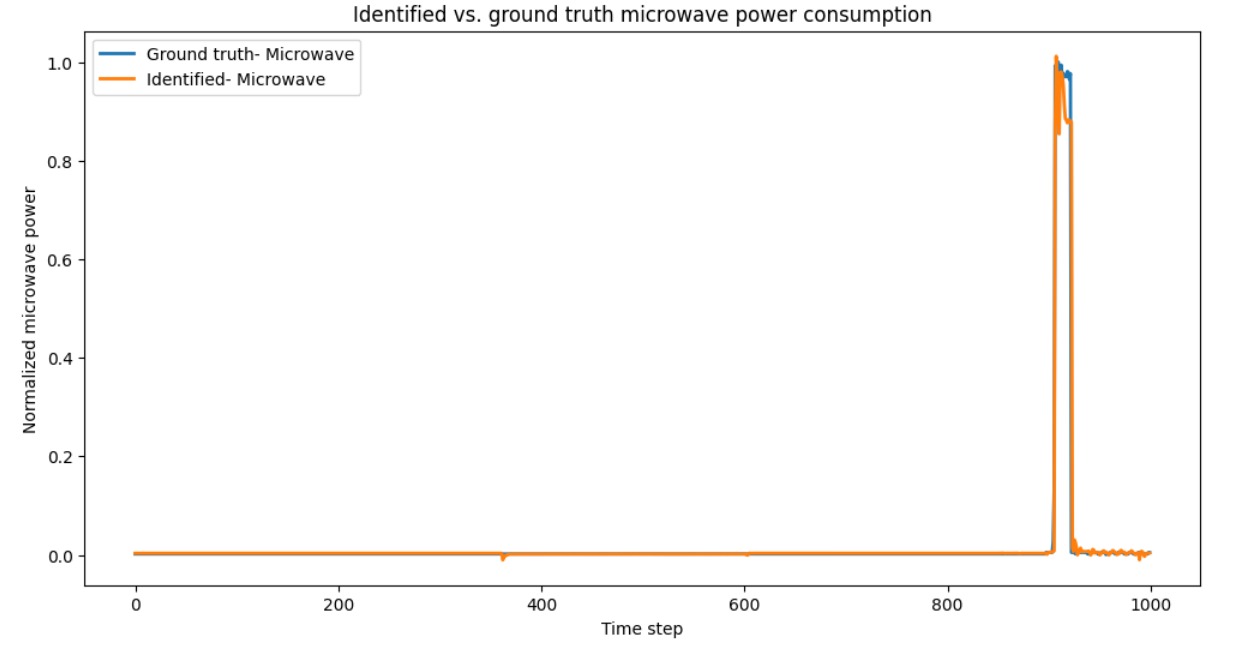}
  \caption{Identified vs. ground truth power consumption for a microwave}
  \label{fig:example}
\end{figure}

\begin{figure}
  \centering
  \includegraphics[width=0.4\textwidth]{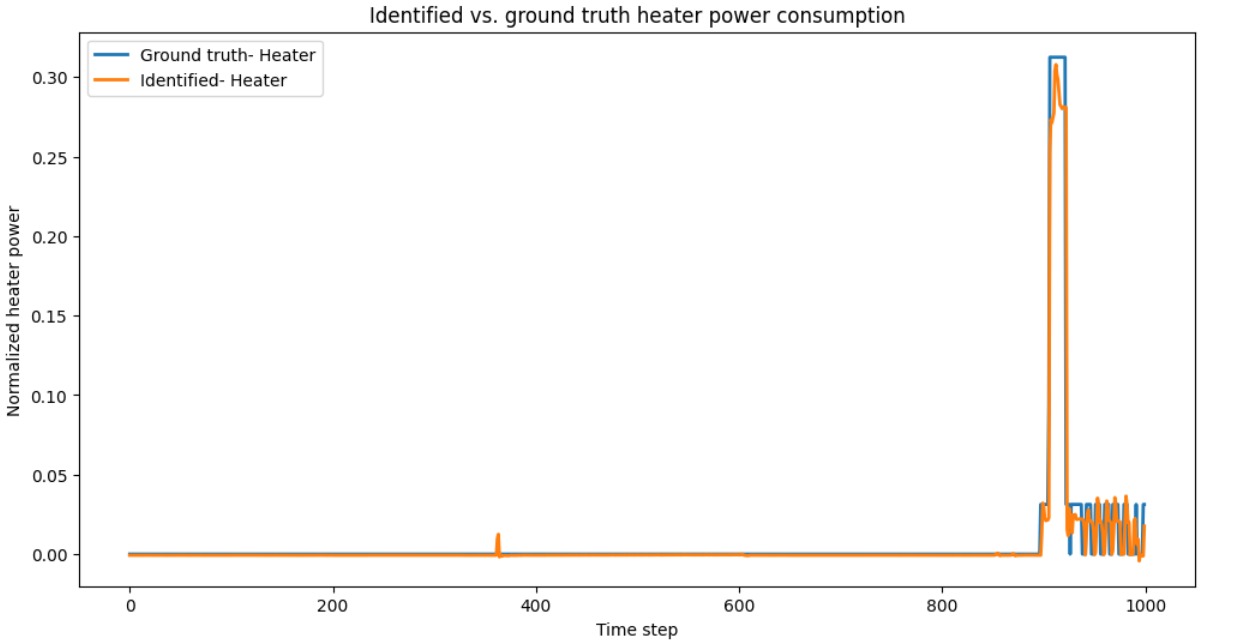}
  \caption{Identified vs. ground truth power consumption for an electric space heater}
  \label{fig:example}
\end{figure}

\begin{figure}
  \centering
  \includegraphics[width=0.4\textwidth]{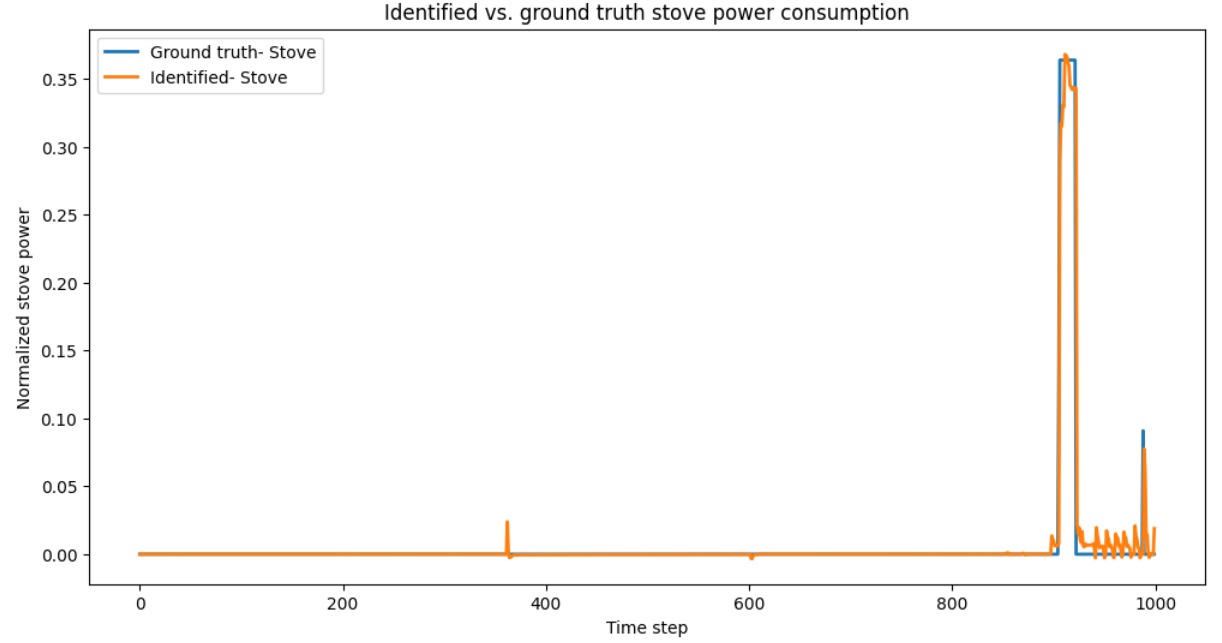}
  \caption{Identified vs. ground truth power consumption for an electric stove}
  \label{fig:example}
\end{figure}

In assessing the performance of our load identification model, we utilize three essential metrics: precision, recall, and F1-score. These metrics are fundamental for evaluating the model's classification accuracy. 
$Precision$ quantifies the accuracy of the model's positive identification of a particular load type as described by \cite{ref27}:
$$
Precision= \frac{True Positives}{True Positives + False Positives} \eqno{(9)}
$$
$Recall$, also known as the true positive rate, measures the model's ability to correctly identify instances of a specific load type among all actual instances of that type, according to the following formula \cite{ref27}:
$$
Recall= \frac{True Positives}{True Positives + False Negatives} \eqno{(10)}
$$
$F1-score$ is defined as the harmonic mean of precision and recall, balancing the trade-off between false positives and false negatives, as follows \cite{ref27}:
$$
F1= \frac{2.Precision.Recall}{Precision + Recall} \eqno{(11)}
$$
Table~II provides a summary of these metrics for test data, illustrating our model's remarkable performance in accurately classifying load types.
\begin{table}
    \centering
    \caption{Model performance}
    \begin{tabular}{|c|c|c|c|}
        \hline
        Appliances & Precision & Recall & F1 \\
        \hline
        dish washer & 1.0000 & 0.9524 & 0.9756\\
        electric space heater & 1.0000 & 1.0000 & 1.0000 \\
        electric stove & 1.0000 & 1.0000 & 1.0000 \\
        fridge & 1.0000 & 0.9890 & 0.9940\\
        microwave & 0.9810 & 1.0000 & 0.9901\\
        washer dryer & 0.9823 & 0.9901 & 0.9860\\
        \hline
    \end{tabular}
    \label{tab:1}
\end{table}

\section{Conclusions}

We propose a CNN-BILSTM model, complemented by an attention mechanism, as a robust solution for the non-intrusive load monitoring (NILM) of problem using low-frequency data. Our proposed model exhibits a remarkable advantage in terms of computational efficiency, making it suitable for real-time applications, and also outperforms existing methods in terms of accuracy. A central strength of our approach lies in its cost-effectiveness because of utilizing low-frequency data. For the future research, one can conduct an in-depth time-correlation analysis within the NILM context to uncover the intricate interplay and dependencies among household appliances to further improve the identification performance.

%\addtolength{\textheight}{-12cm}   % This command serves to balance the column lengths
                                  % on the last page of the document manually. It shortens
                                  % the textheight of the last page by a suitable amount.
                                  % This command does not take effect until the next page
                                  % so it should come on the page before the last. Make
                                  % sure that you do not shorten the textheight too much.

%%%%%%%%%%%%%%%%%%%%%%%%%%%%%%%%%%%%%%%%%%%%%%%%%%%%%%%%%%%%%%%%%%%%%%%%%%%%%%%%

\end{document}